\documentclass[12pt]{iopart}
\usepackage{bm}
\usepackage{graphicx}
\usepackage{cite}
\usepackage{color}
\usepackage[binary,amssymb]{SIunits}
\usepackage{color} 
\usepackage{cancel}

\begin{document}

\title{Two-dimensional MoS$_2$ electromechanical actuators}

\author{Nguyen T. Hung$^1$, Ahmad R. T. Nugraha$^1$, Riichiro Saito$^1$}
\address{$^1$Department of Physics, Tohoku University, Sendai 980-8578, Japan}
\ead{nguyen@flex.phys.tohoku.ac.jp}
\vspace{10pt}
\begin{indented}
\item[]February 2014
\end{indented}

\begin{abstract}
We investigate  electromechanical properties of two-dimensional MoS$_2$ monolayers in the 1H, 1T, and 1T$^\prime$ structures as a function of charge doping by using density functional theory. We find isotropic elastic moduli in the 1H and 1T structures, while the 1T$^\prime$ structure exhibits an anisotropic elastic modulus. Moreover, the 1T structure is shown to have a negative Poisson's ratio, while Poisson's ratios of the 1H and 1T$^\prime$ are positive. By charge doping, the monolayer MoS$_2$ shows a reversibly strain and work density per cycle ranging from $-0.68$\% to $2.67$\% and from 4.4 to 36.9 MJ/m$^3$, respectively, making them suitable for applications in electromechanical actuators. Stress generated is also examined in this work and we find that 1T and 1T$^\prime$ MoS$_2$ monolayers relatively have better performance than 1H MoS$_2$ monolayer. We argue that such excellent electromechanical performance originate from the electrical conductivity of the metallic 1T and semimetallic 1T$^\prime$ structures high Young's modulus of about $150-200$ GPa.
\end{abstract}

\maketitle

\section{Introduction}
Natural muscle is an example of good-performance actuator with work
cycles involving contractions of more than $20$\%, although the stress
generation ability of natural muscle is quite low
($0.35$~MPa)~\cite{madden2004artificial} compared with mechanical
machine.  Various actuation materials have been studied to replace
natural muscle that can directly convert electrical energy into
mechanical energy, with wide potential applications in soft robotics,
adaptive wings for aircraft, and biometric
machines~\cite{madden2004artificial}.  Some well-known actuation
materials, such as carbon nanotubes
(CNTs)~\cite{baughman1999carbon,hung2017charge} and
graphene~\cite{rogers2011graphene,xie2011load}, were shown to generate
larger stress than natural muscle and also larger strain ($\sim1$\%)
than ferroelectric materials ($0.1-0.2$\%) due to their high Young's
moduli of about 1 TPa~\cite{yu2000tensile,hung2016intrinsic}.
Recently, Weissmuller et
al.~\cite{weissmuller2003charge,jin2009nanoporous} showed that Au-Pt
alloys with a network of nanometer-sized pores are good candidates for
the actuation materials because the linear strain reaches $\sim1.3$\%
and work density per cycle is up to 6 MJ/m$^3$, which is a performance
indicator of the muscle.  However, the use of CNTs, graphene, and
Au-Pt nanoporous metal as the electromechanical actuator materials are
still limited mainly because such actuator materials are expensive and
there are difficulties in synthesis, which also make the development
of artificial muscle quite stagnant.  

Very recently, Acerce et al.~\cite{acerce2017metallic} showed a
significant performance on the electromechanical actuation of
two-dimensional metallic molybdenum disulfide (MoS$_2$) nanosheet.
The MoS$_2$ nanosheet is able to generate mechanical stresses of about
$17$ MPa and strains of about $0.8$\%, which leads to the work density
for freely actuated MoS$_2$ films of about $81$ kJ/m$^3$. The MoS$_2$
nanosheet actuator is also able to lift more than 150 times its own
weight at low voltages $\pm0.3$ V for hundreds of cycles. High
actuation performance of the 1T MoS$_2$ nanosheet originates from the
high electrical conductivity of the metallic 1T structure and their
elastic modulus of $2.5$ GPa.  However, their study is limited to the
1T MoS$_2$ nanosheet (or the multilayer
MoS$_2$) although it is known that MoS$_2$ could have at
least three different stable forms that have been synthesized so far:
1H, 1T, and 1T$^\prime$
phases~\cite{lin2014atomic,naylor2016monolayer}.  Furthermore, the
condition of charge doping that can possibly support for high
actuation performance of the \emph{monolayer} MoS$_2$ layers is still unclear in both experimental and theoretical investigations.

With the above backgrounds, it is highly desirable to explore the
strain, stress, work density and electronic structure of two-dimensional MoS$_2$ under
the charge doping to understand the best conditions or the best
structures for electromechanical actuator, which can be evaluated by
first-principles calculations.  Theoretically, it is expected that
\emph{monolayer} MoS$_2$ has higher electrical conductivity and larger
surface area than the MoS$_2$ nanosheet ~\cite{li2012ideal,bertolazzi2011stretching,castellanos2012elastic}.
Therefore, in this work we will focus our attention on the
electromechanical actuator performance of the 1H, 1T, and 1T$^\prime$
MoS$_2$ monolayers as a function of charge doping for both electron
and hole doping.  As the main highlight of this paper, our calculated
results reveal that the 1T and 1T$^\prime$ MoS$_2$ monolayers
relatively have better performance and better actuator response than
1H MoS$_2$ monolayer.  In addition, depending on the structure, we can
have either isotropic or anisotropic actuation properties in the
MoS$_2$ monolayers.

\section{Method}
\subsection{Calculation details}

\begin{figure}[t!]
  \centering \includegraphics[clip,width=10cm]{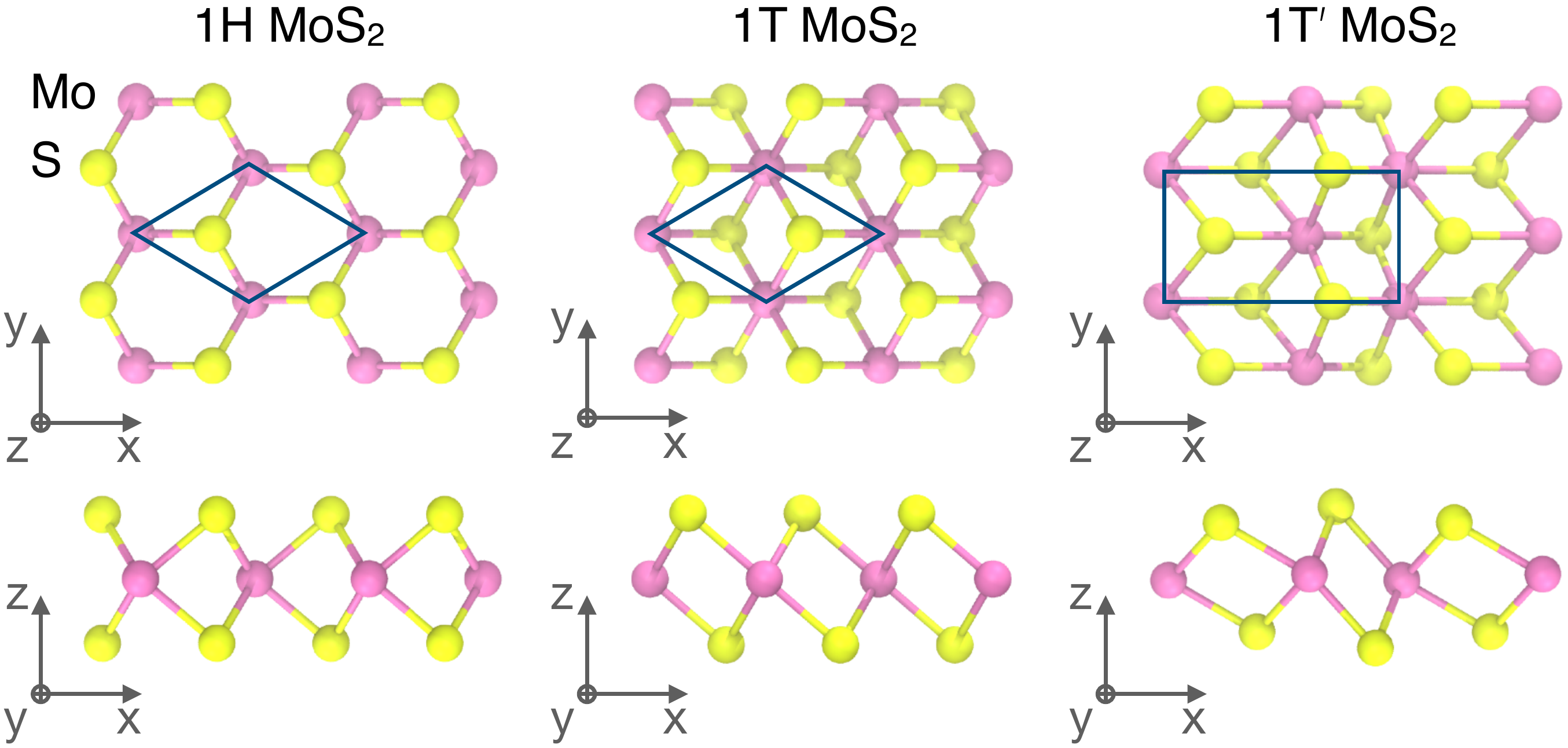}
  \caption{\label{fig:model} Top and side views of 1H, 1T, and
    1T$^\prime$ structures of monolayer MoS$_2$.  The 1H, 1T, and
    1T$^\prime$ structures are composed of trigonal, octahedral and
    distorted octahedral lattices, respectively.}
\end{figure}

In Fig.~\ref{fig:model}, we show the 1H, 1T, and 1T$^\prime$
structures of the monolayer MoS$_2$.  The 1H MoS$_2$ structure is
based on trigonal lattice, where the S atoms are located in a
hexagonal close-packed structure while the Mo atoms are sandwiched
between two atomic layers of S atoms in a trigonal prismatic geometry.
In the cases of 1T MoS$_2$ and 1T$^\prime$ MoS$_2$, the Mo atoms are
octahedrically ordered and disordered, respectively, surrounded by the
S atoms.  The primitive unit cells of the 1H and 1T MoS$_2$ are
hexagonal with the optimized lattice parameters of $3.19$ \AA\ and
$3.18$ \AA, respectively, while the unit cell of the 1T$^\prime$
MoS$_2$ is rectangular with the optimized lattice parameters of
$a=5.72$ \AA\ and $b=3.16$ \AA\ as shown in
Fig.~\ref{fig:model}. These lattice parameters are consistent with
previous theoretical results~\cite{sun2016origin,fan2014site}. Since
periodic boundary condition is applied in all models, a vacuum space
of 30 \AA\ in the direction perpendicular to the monolayer ($z$
direction) is used in order to avoid virtual interactions between
layers.

We perform first-principles calculations to determine the total energy
and the electronic structure of monolayer MoS$_2$ using the Quantum
ESPRESSO package~\cite{giannozzi2009quantum-short}.  We use
pseudopotentials from the Standard Solid-State Pseudopotentials
library (accuracy version)~\cite{lejaeghere2016reproducibility-short}.
The exchange-correlation energy is evaluated by the general-gradient
approximation using the using the Perdew-Burke-Ernzerhof
functional~\cite{perdew1996generalized}.  An energy cut-off of 60 Ry
is chosen for the expansion of the plane waves, which is sufficient to
obtain convergence of total energy.  In our simulation, the
\textbf{k}-point grids in the Brillouin-zone are employed according to
the Monkhorst-Pack scheme~\cite{monkhorst1976special}, where
\textbf{k} is the electron wave vector.  We use $16\times16\times1$,
$16\times16\times1$, and $8\times16\times1$ \textbf{k}-points for the
1H, 1T and 1T$^\prime$ MoS$_2$, respectively.  To obtain optimized
atomic configurations of MoS$_2$ monolayers, the atomic positions and
cell vectors are fully relaxed using the
Broyden-Fretcher-Goldfarb-Shanno minimization
method~\cite{broyden1970convergence,fletcher1970new,goldfarb1970family,shanno1970conditioning}
until all the Hellmann-Feynman forces and all components of the stress
are less than $5\times10^{-4}$ Ry/a.u. and $5\times10^{-2}$ GPa,
respectively.

To discuss the electromechanical
actuation of the MoS$_2$ monolayers, the geometry optimization is then
performed for each charge doping from $-0.1$ to $+0.1$ electron per
atom ($e$/atom), in which the electron (hole) doping is simulated by adding
(removing) electrons to the unit cell with the same amount of
uniformly positive (negative) charge in the background so as to keep
the charge neutrality.

\subsection{In-plane mechanical moduli}
In order to obtain mechanical moduli of MoS$_2$ monolayers, we firstly calculate elastic constants $C_{ij}$, which are derived from the finite difference approach by using the Thermo-pw code~\cite{dal2016elastic}.  From the point of view of elasticity theory, it is known that the values of $C_{ij}$ are related to the equivalent volume of the unit cell.  Because a vacuum
space is left along the $z$ direction in the unit cell, the calculated
$C_{ij}$ must be rescaled by $h/d_0$, where $h$ is the length of the
cell along $z$ axis and $d_0$ is the effective layer thickness of the
monolayer MoS$_2$.  In the present study, we set $d_0=6.145$ \AA\,
i.e., one half of the out-of-plane lattice constant of bulk
MoS$_2$~\cite{young1968lattice}. The angular dependence of the in-plane ($xy$-plane) Young's modulus $Y(\theta)$ and Poisson's ratio $\nu(\theta)$ are then expressed as~\cite{wang2015electro}
\begin{equation}
\label{eq:young}
  Y(\theta)=\frac{C_{11}C_{22}-C_{12}^2}{C_{11}\alpha^4+C_{22}\beta^4-
  \left(2C_{12}-\displaystyle\frac{C_{11}C_{22}-C_{12}^2}{C_{66}}\right)
  \alpha^2\beta^2},
\end{equation}
and
\begin{equation}
\label{eq:poisson}
  \nu(\theta)=\frac{C_{12}(\alpha^4+\beta^4)-\left(C_{11}+C_{22}-
  \displaystyle\frac{C_{11}C_{22}-C_{12}^2}{C_{66}}\right)
  \alpha^2\beta^2}{C_{11}\alpha^4+C_{22}\beta^4-
  \left(2C_{12}-\displaystyle\frac{C_{11}C_{22}-C_{12}^2}{C_{66}}\right)
  \alpha^2 \beta^2},
\end{equation}
where $\theta$ is the angle relative to the $x$ direction, $\alpha=\sin(\theta)$, $\beta=\cos(\theta)$, and $C_{ij}$ are
the elastic constants obtained from the first-principles calculations.
Since monolayer MoS$_2$ is a two-dimensional structure, there are
four independent elastic constants $C_{11}$, $C_{22}$, $C_{12}$, and
$C_{66}$.

\subsection{Work-per-cycle analysis}
\begin{figure}[t!]
  \centering \includegraphics[clip,width=6cm]{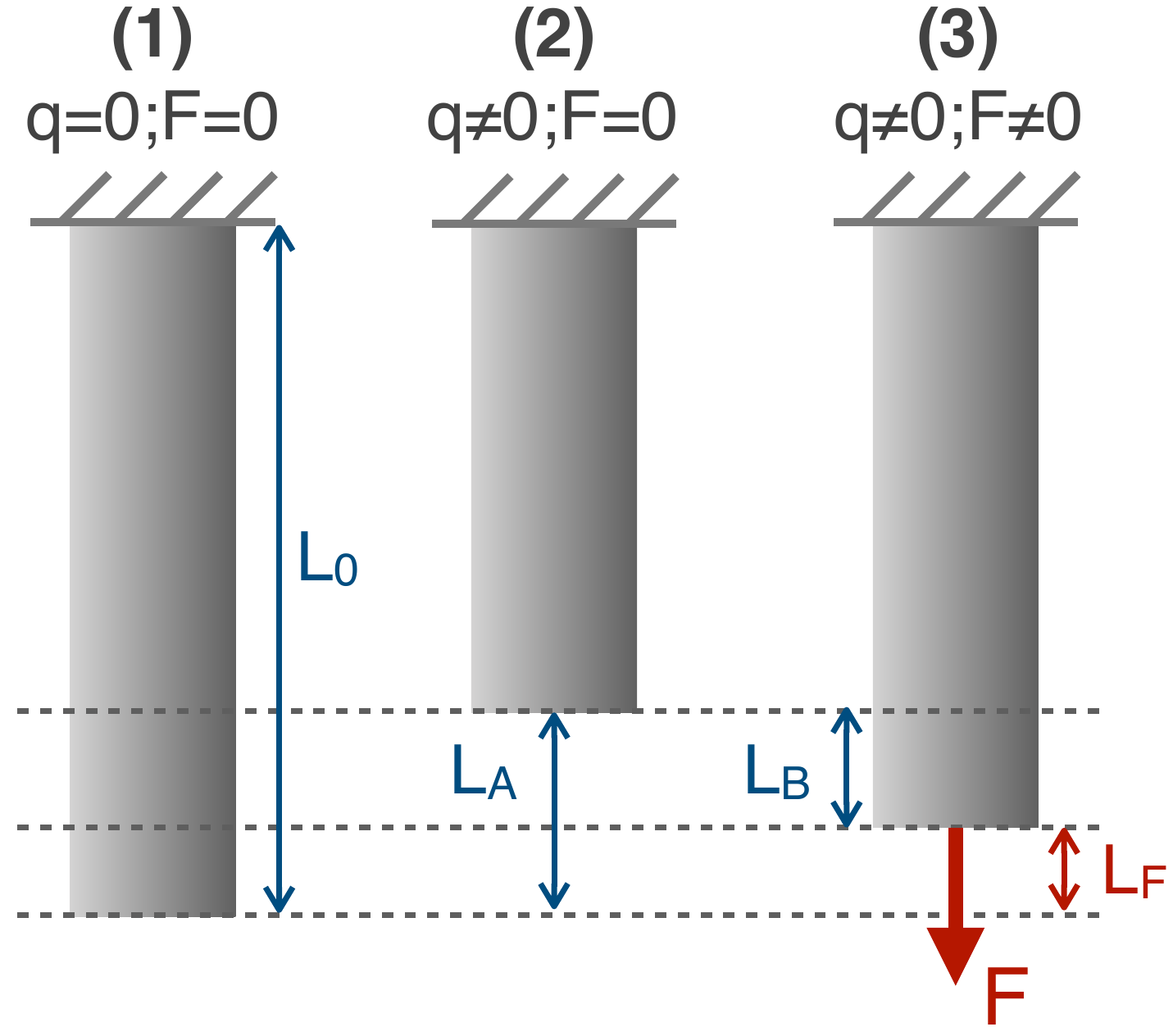}
  \caption{\label{fig:work-cycle}Length changes occurring during the
    loading and actuation processes.}
\end{figure}

As an important performance indicator for actuation, we adopt the
work density per cycle $W$ of an actuator
for discussion of the monolayer MoS$_2$.  We assume that the actuator is a linear
elastic solid and the general condition to be considered is
illustrated in Fig.~\ref{fig:work-cycle}.  There are three states when
the actuator is loaded by a constant tensile force: (1) the material
is at zero charge doping $q=0$ with an initial length $L_0$, (2)
applying a charge doping $q\neq 0$ produces a length change $L_A$ due
to the electromechanical actuation process, and (3) applying a force
$F\neq 0$ produces a deformation $L_B$. According to Hooke's law,
which is generally true at small strains, $L_A$ and $L_B$ are given by
\begin{equation}
\label{eq:length}
L_A=\epsilon L_0,\hspace{1em} L_B=\frac{FL_0}{AY},
\end{equation}
where $\epsilon$, $Y$, and $A$ are, respectively, the strain, Young's modulus, and
cross-sectional area of monolayer MoS$_2$ after the charge doping has
been applied.  The work density per cycle that includes steps (1)-(3) as
shown in Fig.~\ref{fig:work-cycle} is given by~\cite{spinks2005work}
\begin{equation}
\label{eq:work}
W=\frac{F(L_A-L_B)}{V},
\end{equation}
where $V=L_0A$ is the volume.  By substituting $L_A$ and $L_B$ in
Eq.\ref{eq:length} into Eq.\ref{eq:work}, $W$ can be written as
\begin{equation}
\label{eq:work-density}
W=\frac{F\epsilon}{A}-\frac{F^2}{A^2Y}.
\end{equation}
We can determine the maximum $W$ from Eq.\ref{eq:work-density} by
solving $\mathrm{d}W/\mathrm{d}F=0$.  The formula for the
maximum work density per cycle is given by
\begin{equation}
\label{eq:work-density-1}
W_{\mathrm{max}}=\frac{1}{4}Y\epsilon^2,
\end{equation}
when $F_{\mathrm{max}}=\frac{1}{2}Y\epsilon A$. However, in most of
experiments, the work density is often expressed in terms of stored
energy density $W_s$, which is calculated from the linear relation between
$\sigma$ and $\epsilon$, giving the formula
\begin{equation}
\label{eq:work-density-2}
  W_s =\frac{1}{2}Y\epsilon^2 = 2 W_{\mathrm{max}}.
\end{equation}
Equation~\ref{eq:work-density-2} will be used to compare our
theoretical results with recent experimental data of MoS$_2$
electromechanical actuators~\cite{acerce2017metallic}.

\section{Results}

\subsection{Mechanical properties}

\begin{figure}[t!]
  \centering \includegraphics[clip,width=8.5cm]{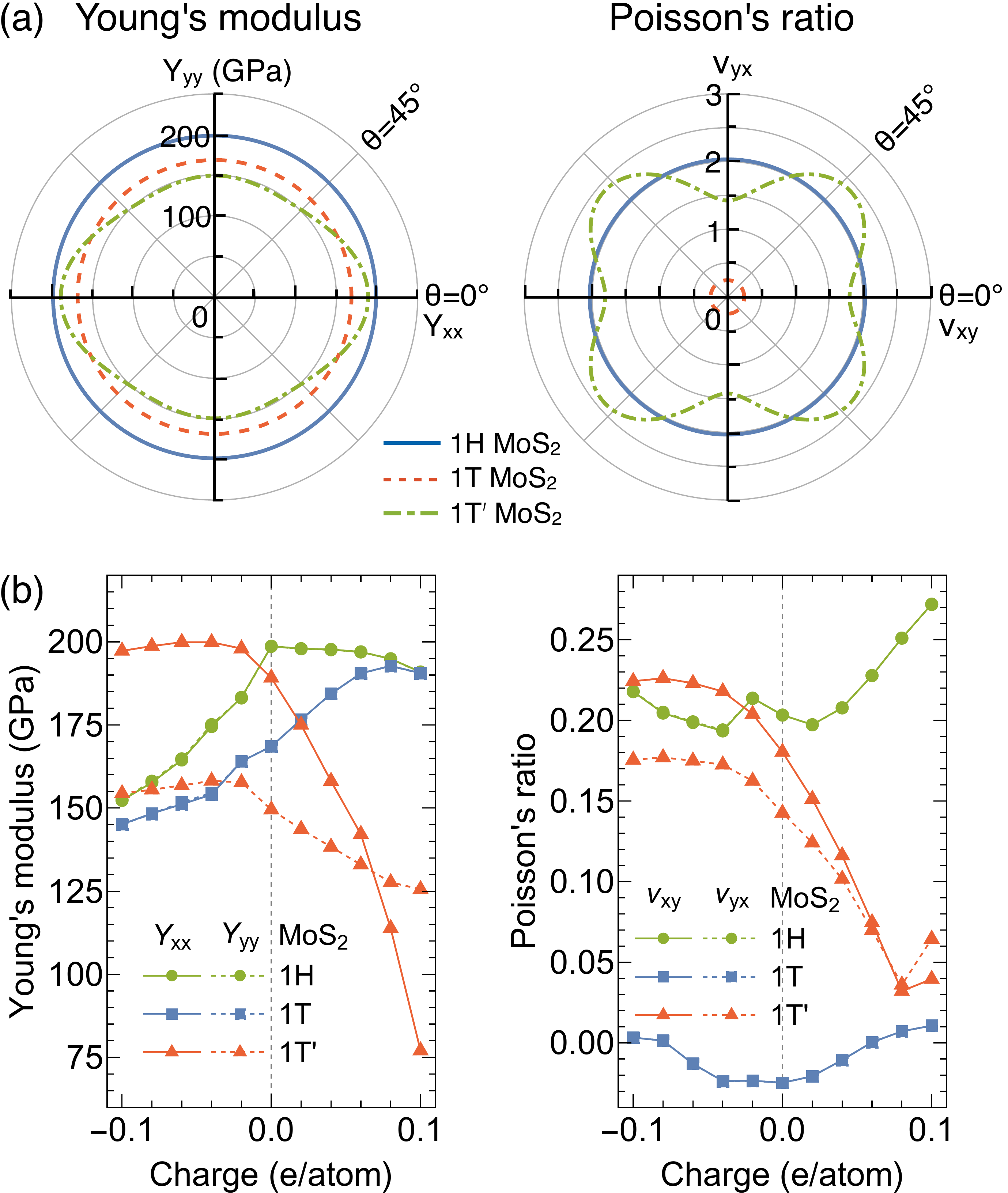}
  \caption{\label{fig:young-poisson}(a) Polar diagram for Young's modulus
    $Y$ (left) and Poisson's ratio $\nu$ (right) of monolayer MoS$_2$
    with 1H, 1T, and 1T$^\prime$ structures.  The angle $\theta$
    identifies the direction of applied force with respect to the
    $x$-axis.  Isotropic (anisotropic) behaviour is associated with a
    circular (noncircular) shape of the polar plot. (b) Young's modulus and
    Poisson's ratio of monolayer MoS$_2$ plotted as function of charge
    doping per atom.}
\end{figure}

\begin{table}[t]
  \caption{Elastic constants $C_{ij}$ (GPa), Young's modulus $Y$ (GPa)
    and Poisson's ratio $\nu$ of the monolayer MoS$_2$ with 1H, 1T,
    and 1T$^\prime$ structures.}
\centering  
\begin{tabular}{l l l l l l l l l}
\hline
MoS$_2$ & $C_{11}$& $C_{22}$& $C_{12}$& $C_{66}$& $Y_{xx}$& $Y_{yy}$& $\nu_{xy}$& $\nu_{yx}$\\ 
\hline
1H & 207& 207& 42& 83& 197& 197& 0.20& 0.20\\  
1T  & 169& 169& -4& 86& 167& 167& -0.02& -0.02\\
1T$^\prime$ & 194& 153&  28& 61& 189& 150& 0.18& 0.14
\end{tabular}
\label{table:elastic}    
\end{table}

To discuss the actuator response of the monolayer MoS$_2$, we firstly
check the mechanical moduli at the neutral condition and at the charge
doping cases. In Fig.~\ref{fig:young-poisson}a, we show the dependences of $Y$ and $\nu$ on the direction of monolayer MoS$_2$ at the neutral condition,
i.e. $q=0$. The shape of $Y$ and $\nu$ in the polar plot indicates not
only the elastic isotropy in the 1H and 1T MoS$_2$ monolayers, but
also the elastic anisotropy in the 1T$^\prime$ MoS$_2$ monolayer.  The
anisotropy of elastic moduli in the 1T$^\prime$ MoS$_2$ monolayer
originates from the fact that the low-symmetry 1T$^\prime$ structure
is a distorted one from the high-symmetry 1T structure.  For
comparison, the values of $C_{ij}$, $Y$, and $\nu$ of the 1H, 1T and
1T$^\prime$ MoS$_2$ structures are listed in
Table~\ref{table:elastic}.  We can see that $Y$ is found to be $199$
GPa for the 1H MoS$_2$, which is in a good agreement with a previous
theoretical result ($Y=200$ GPa)~\cite{li2012ideal}.  Bertolazzi et
al.~\cite{bertolazzi2011stretching} obtained an effective $Y$ of
$270\pm100$ GPa for the monolayer MoS$_2$, while Castellanos-Gomez et
al.~\cite{castellanos2012elastic} obtained an average $Y$ of $210-370$
GPa for multilayer MoS$_2$ consisting of 5 to 25 layers.  Note that
both
experiments~\cite{bertolazzi2011stretching,castellanos2012elastic}
using atomic force microscope tip applied on the monolayer or
multilayer MoS$_2$ suspended on the substrate containing an array of
circular holes are under biaxial tensile stress.  Therefore, the
experimental results of biaxial elastic modulus are higher than the
theoretical results of uniaxial elastic modulus in this present study.
We obtained the Young modulus $Y=167$ GPa for the 1T MoS$_2$, while
$Y_{xx}=189$ GPa and $Y_{yy}=150$ GPa for the 1T$^\prime$ MoS$_2$.

The values of Poisson's ratio of the monolayer MoS$_2$ are unique
because the 1T structure exhibits a negative Poisson's ratio of
$-0.02$, while Poisson's ratios of the 1H ($\nu=0.20$) and 1T$^\prime$
MoS$_2$($\nu_{xy}=0.18$ and $\nu_{yx}=0.14$) are positive.  When a
compressive (tensile) strain is acted in one direction, materials tend
to expand (contract) in the perpendicular direction, corresponding to
positive Poisson's ratio for ordinary materials.  The opposite is the
situation for materials with negative Poission's ratio.  We note that
beside our study, there have been reports that Poisson's ratio can be
negative in other 2D materials such as in black phosphorus
($\nu=-0.5$)~\cite{du2016auxetic}, single-layer graphene ribbons
($\nu=-1.51$)~\cite{jiang2016negative}, and $\delta$-phosphorene
($\nu=-0.267$)~\cite{wang2017delta}.  We expect that exploring 2D
materials with negative Poission's ratio could have useful
applications, for example, as vanes for aircraft gas turbine engines,
sponges, and fasteners~\cite{baughman1998negative}.




In Fig.~\ref{fig:young-poisson}b, we show $Y$ and $\nu$ of the
monolayer MoS$_2$ as a function of charge doping, i.e. $q\ne 0$.  For
the electron doping ($q<0$), $Y$ of the 1H and 1T structures is
decreased, while $Y$ of the 1T$^\prime$ structure is increased.  For
the hole doping ($q>0$ $e$/atom), $Y$ of 1T$^\prime$ structure is
increased, while $Y$ of the 1H and 1T$^\prime$ structures decreased.
The maximum $Y$ of monolayer MoS$_2$ of about $200$ GPa is smaller
than that of carbon-based structures ($400-1000$
GPa)~\cite{hung2017charge,hung2017three}, but is comparable to that of
stainless steel (192 GPa)~\cite{rho1993young}.  The high $Y$ values of
monolayer MoS$_2$ are important for artificial muscle applications
since it could generate large force per unit area. Moreover, a
significant change of $\nu$ is found in the 1T$^\prime$ structure from
$0.23$ to $0.03$, as shown in Fig.~\ref{fig:young-poisson}b.  For the
1H and 1T structures, $\nu$ increases with increasing $|q|$ for both
electron and hole doping.  It should be noted that Poisson's ratio of
the 1T structure becomes positive at $q=-0.08$ $e$/atom ($\nu=0$) and
at $q=0.06$ $e$/atom ($\nu=0$) for the electron and hole doping,
respectively.

\subsection{Actuator response}

\begin{figure}[t!]
  \centering \includegraphics[clip,width=7cm]{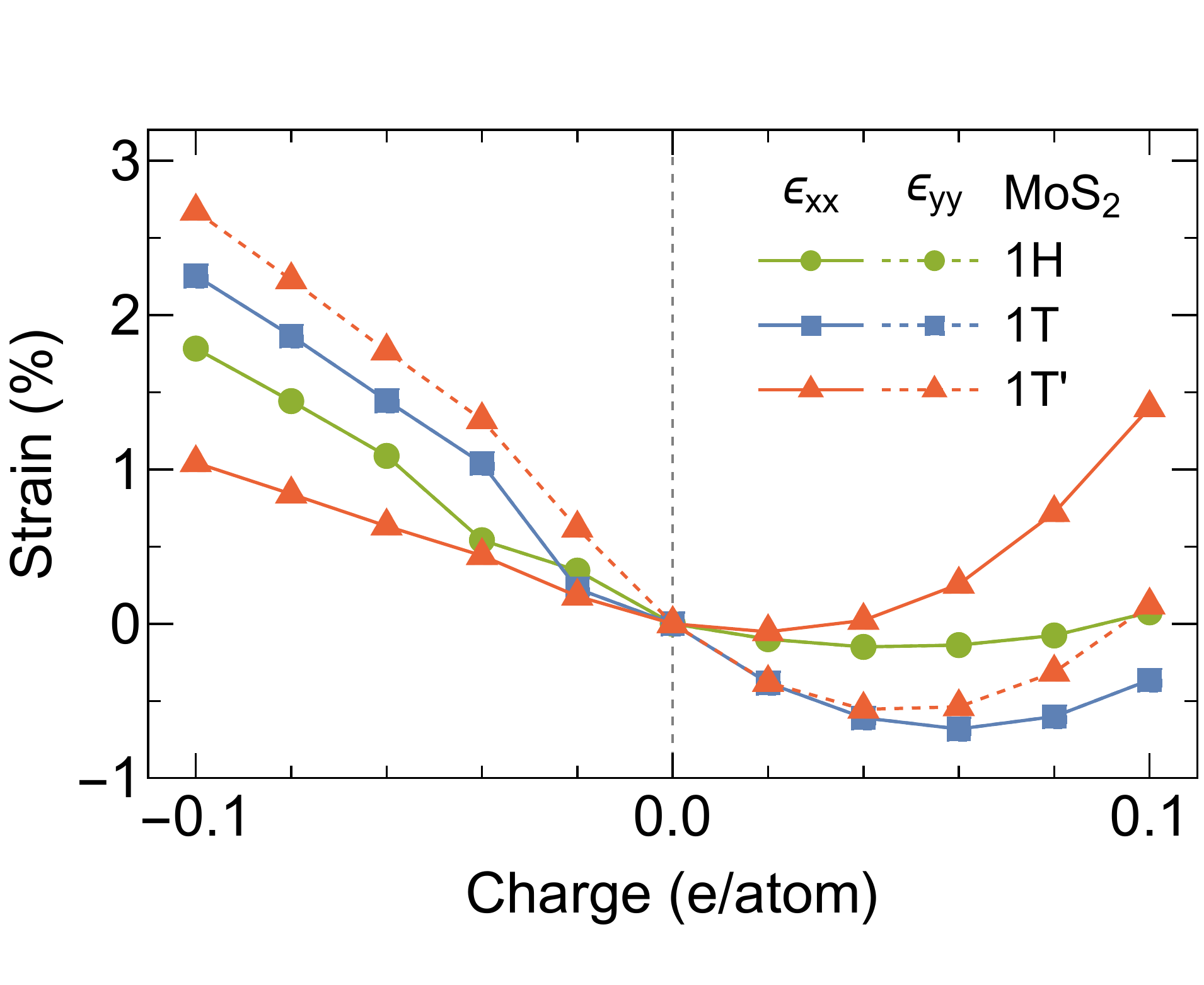}
  \caption{\label{fig:energy-strain}Strain as function of charge doping per
    atom of monolayer MoS$_2$.}
\end{figure}

In order to study the variation of the structural deformation as a
function of charge doping, we define the in-plane strain as
\begin{equation}
\label{eq:strain}
  \epsilon_{xx}=\Delta a/a_0, \quad \epsilon_{yy}=\Delta b/b_0,
\end{equation}
where $a_0$ and $b_0$ are, respectively, the length of the unit cell
in $x$ and $y$ directions at geometry optimization for neutral case,
and $\Delta a$ and $\Delta b$ are the increment (or decrement) of
$a_0$ and $b_0$, respectively, after the charge doping has been
applied. In Fig.~\ref{fig:energy-strain}, we show the strain for each
monolayer MoS$_2$ as a function of charge doping $q$ ranging from
$-0.1$ to $0.1$ $e$/atom.  This charge range is reasonable because a
typically accessible charge in experiments is ranging from $-0.3$ to
$0.1$ $e$/atom~\cite{sun2002dimensional}.  In the neutral case, we
obtain strains $\epsilon_{xx}=\epsilon_{yy}=0$.  For the electron
doping, i.e. $q<0$, $\epsilon_{xx}$ and $\epsilon_{yy}$ are
approximately a linear function of $q$.  At $q=-0.1$ $e$/atom, the
strains of the 1H and 1T MoS$_2$ are up to $1.78$\% and $2.25$\%,
respectively.  We can also say that 1H and 1T MoS$_2$ monolayers will
expand isotropically ($\epsilon_{xx} = \epsilon_{yy}$).  On the other
hand, the 1T$^\prime$ MoS$_2$ shows an anisotropic expansion with
$\epsilon_{xx}=1.04$\% and $\epsilon_{yy}=2.67$\%.  For the hole
doping, i.e. $q>0$, $\epsilon_{xx}$ and $\epsilon_{yy}$ are a
non-linear function of $q$.  The 1H and 1T MoS$_2$ monolayers show an
isotropic compression with the maximum strains of about $-0.15$\% and
$-0.68$\% at $q=0.04$ $e$/atom and $0.06$ $e$/atom, respectively, as
shown in Fig.~\ref{fig:energy-strain}.  On the contrary, the
1T$^\prime$ MoS$_2$ shows an anisotropic behaviour with the expansion
strain ($\epsilon_{xx}=1.40$\% at $q=0.1$ $e$/atom) and the
compression strain ($\epsilon_{yy}=-0.55$\% at $q=0.04$ $e$/atom)
along $x$ and $y$ directions, respectively.  In this present study,
the strain magnitude of $0.68$\% of the 1T MoS$_2$ by the hole doping
is in a good agreement with the experimental data of about
$0.6-0.8$\%~\cite{acerce2017metallic} that making them suitable for
applications in electromechanical actuators.

\subsection{Actuator performance}

\begin{figure}[t!]
  \centering \includegraphics[clip,width=8cm]{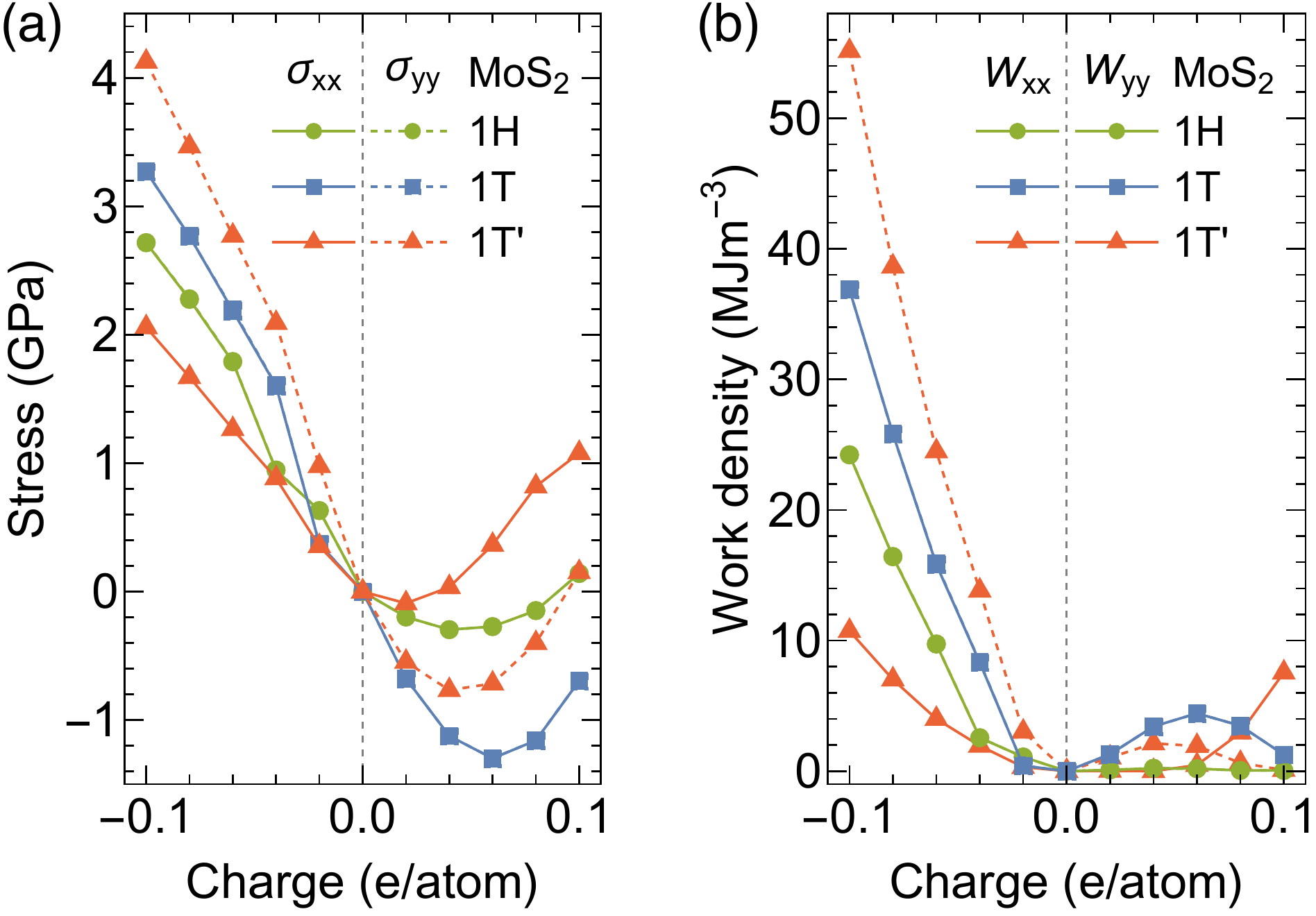}
  \caption{\label{fig:stress}(a) Work density and (b) stress generated
    by monolayer MoS$_2$ plotted as function of charge (electron and
    hole) doping per atom.}
\end{figure}

The power of electromechanical actuators is characterized by the
stress generated, which is determined by the product of the elastic
modulus and the strain: $\sigma=Y\epsilon$.  In
Fig.~\ref{fig:stress}b, we show the stress generated in monolayer
MoS$_2$ as a function of charge doping. In the neutral case, we obtain
$\sigma_{xx}=\sigma_{yy}=0$ because $\epsilon_{xx}=\epsilon_{yy}=0$.
For the electron doping at $q=-0.1$ $e$/atom, we obtain
$\sigma_{xx}=\sigma_{yy}=2.72$ GPa and $3.27$ GPa for the 1H and 1T
structures, respectively, while $\sigma_{xx}=2.06$ GPa and
$\sigma_{yy}=4.13$ GPa for the 1T$^\prime$ structure.  For the hole
doping, the maximum stress ($\sigma_{xx}=\sigma_{yy}=-1.30$ GPa) is
found in the 1T structure at $q=0.06$ $e$/atom.  Our calculated
$\sigma$ ranging from $-1.30$ to $3.27$ GPa for the 1T MoS$_2$
monolayer is higher than the experimental value for the 1T MoS$_2$
nanosheet ($0.017$ GPa)~\cite{acerce2017metallic} due to its high
Young's modulus (see Fig.~\ref{fig:young-poisson}b) and it is
comparable to the carbon nanotube actuators ($\sim3$
GPa)~\cite{qu2008carbon,hung2017charge}. Our results suggest that the
electron doping should be good for the actuator application of MoS$_2$
monolayers.

The performance of electromechanical actuators is characterized by the
work density per cycle that is defined Eq.~\ref{eq:work-density-2}
as $W=Y\epsilon^2/2$. In Fig.~\ref{fig:stress}a, we calculate $W$ of
the monolayer MoS$_2$ as a function of charge doping.  For the 1T
MoS$_2$, $W$ is up to 36.9 MJ/m$^{3}$ at $q=-0.1$ $e$/atom and 4.4
MJ/m$^{3}$ at $q=0.06$ $e$/atom for the electron and hole dopings,
respectively, which is more than 100-1000 times that of skeleton
muscle ($~\sim 0.04$ MJ/m$^{3}$)~\cite{madden2004artificial}.  These
results are much higher than the experimental values for the 1T
MoS$_2$ nanosheet ($0.081$ MJ/m$^{3}$)~\cite{acerce2017metallic} since
$Y$ ($145-193$ GPa) of the monolayer MoS$_2$ is larger than that of
the MoS$_2$ nanosheet ($Y=2.5\pm0.1$ GPa)~\cite{acerce2017metallic}.
For the 1H and 1T$^\prime$ structures, $W$ at the electron doping case
is higher than that of the hole doping case, which suggest that the
electron doping should be good to achieve high-performance
electromechanical actuators.

\subsection{Electronic properties}

\begin{figure}[t!]
  \centering \includegraphics[clip,width=8.5cm]{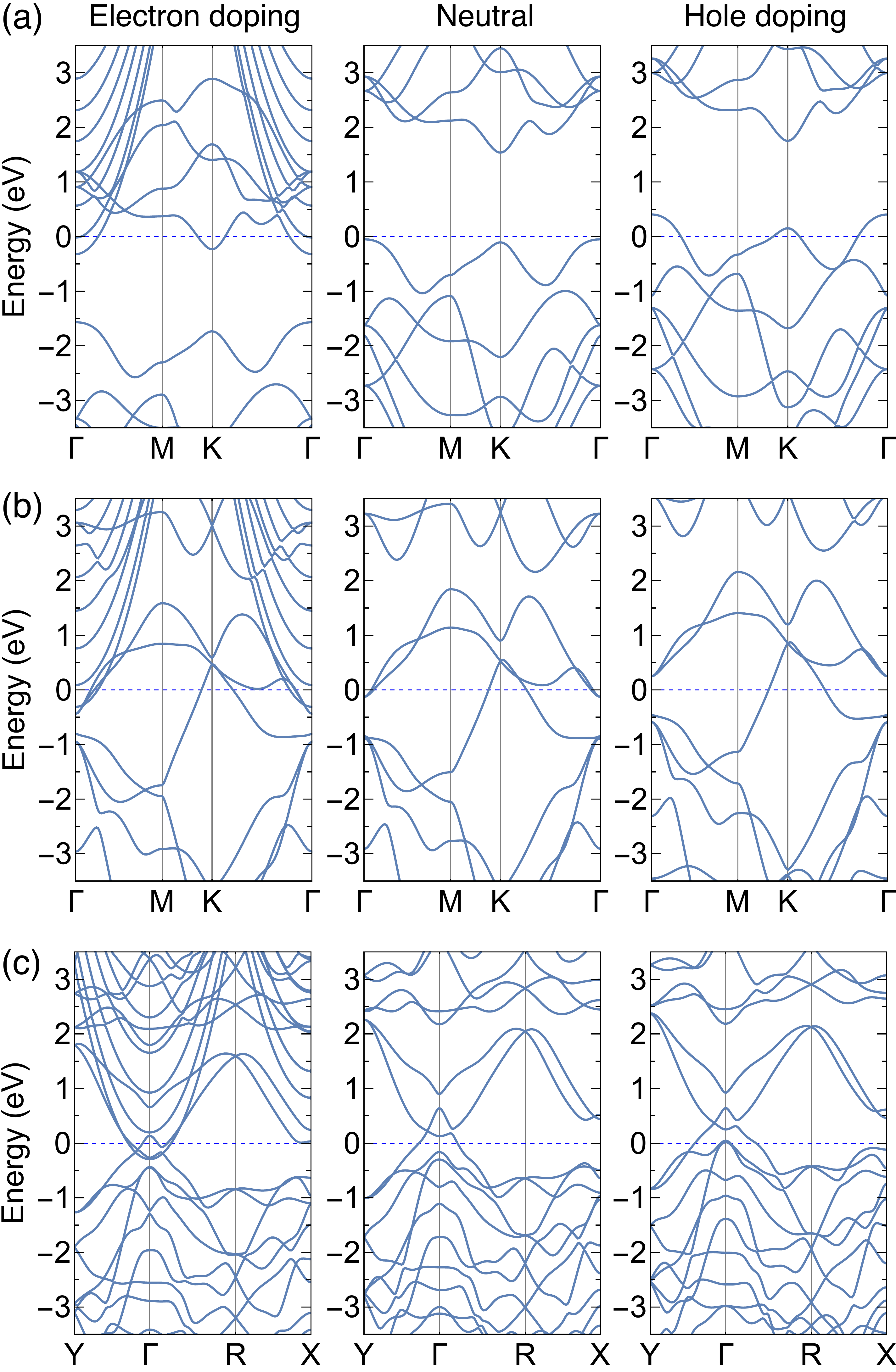}
  \caption{\label{fig:bands}Energy band structures of (a) 1H, (b) 1T,
    and (c) 1T$^\prime$ MoS$_2$ monolayers with different electron
    doping ($q=-0.1$ $e$/atom) and hole doping ($q=+0.1$ $e$/atom)
    including those in the neutral condition ($q=0$ $e$/atom).  The
    Fermi energy (dashed line) is set to zero for all plots.}
\end{figure}

To understand the variation of the electronic properties of the
monolayer MoS$_2$ under charge doping, finally, we can examine the energy band
structures of the monolayer MoS$_2$ within the range of charge doping
considered in the present work. In Figs.~\ref{fig:bands}a-c, we show,
respectively, the calculated electronic structures of the 1H, 1T and
1T$^\prime$ MoS$_2$ along the high-symmetry points of their
corresponding Brillouin zone for neutral and charge doping states.
From Fig.~\ref{fig:bands}a, we can see that in the neutral case, the
1H MoS$_2$ monolayer is an indirect-gap semiconductor (the top of
valence band is at the $\Gamma$ point while the bottom of conduction
band is at the $K$ point) with the band gap of about 1.59 eV.  The
electron (hole) doping does (does not) transform the 1H MoS$_2$
monolayer to be a direct-gap semiconductor.  The direct (indirect)
band-gap of 1H MoS$_2$ monolayer in the case of electron (hole) doping
is about 1.25 (1.35) eV.  On the other hand, from
Figs.~\ref{fig:bands}b and~c, we find that basically in both cases of
charge doping and neutral condition, the 1T MoS$_2$ is a metal, while
the 1T$^\prime$ MoS$_2$ is a semimetal, with an exception that the
1T$^\prime$ MoS$_2$ transforms to a metal by heavy electron doping.  A
common interesting feature we can see in Figs.~\ref{fig:bands}a-c is
that the electron doping ``pull down'' many interlayer bands of the
1H, 1T and 1T$^\prime$ MoS$_2$, while hole doping do not.  Such a
phenomenon might contribute to the higher performance of MoS$_2$
electromechanical actuators by the electron doping rather than the
hole doping, as shown previously in Figs.~\ref{fig:stress}a and~b.

\section{Conclusions}
We have performed a first principles theoretical study on the actuator
performance and on the electronic structure as a function of charge
doping for the 1H, 1T and 1T$^\prime$ MoS$_2$ monolayers. We find that
the work density per cycle and stress generated in 1T and 1T$^\prime$
MoS$_2$ monolayers are relatively larger than those in 1H MoS$_2$
monolayer. This excellent electromechanical performance originate from
the electrical conductivity of the metallic 1T and semimetallic
1T$^\prime$ structures high Young's modulus of about 150--200 GPa
under charge doping.  The results obtained also reveal that the 1H and
1T MoS$_2$ show the actuator isotropy, while 1T$^\prime$ MoS$_2$ shows
the actuator anisotropy, which implies that researchers can have more
freedom to choose the best MoS2$_2$ structures depending on the
isotropic or anisotropic electromechanical applications.

\section{Acknowledgements}
N.T.H. and A.R.T.N acknowledge the Interdepartmental Doctoral Degree
Program for Multidimensional Materials Science Leaders under the
Leading Graduate School Program in Tohoku
University. R.S. acknowledges JSPS KAKENHI Grant Numbers JP25107005
and JP25286005.

\bibliographystyle{iopart-num.bst}

\end{document}